\documentclass{article}

\PassOptionsToPackage{numbers,compress}{natbib}

\usepackage[main,final]{neurips_2026} 

\usepackage[utf8]{inputenc}
\usepackage[T1]{fontenc}
\usepackage{hyperref}
\usepackage{url}
\usepackage{booktabs}
\usepackage{graphicx}
\usepackage{subcaption}
\usepackage{longtable}
\usepackage{float}
\usepackage{placeins}
\usepackage{amsfonts}
\usepackage{amsmath}
\usepackage{nicefrac}
\usepackage{microtype}
\usepackage{xcolor}
\usepackage{colortbl}
\colorlet{gsblue}{blue!10}
\newcommand{\gc}[1]{\cellcolor{gsblue}#1}
\usepackage{rotating}
\usepackage{multirow}
\usepackage{tabularx}
\usepackage{adjustbox}

\title{GS-Codec: A Gaussian-Splatting Bottleneck for Neural Audio Coding}

\author{Ron Aluf \quad Alon Canfi \quad Eliya Nachmani \\
  School of Electrical and Computer Engineering \\
  Ben-Gurion University of the Negev \\
  \texttt{alufr@post.bgu.ac.il, canfia@post.bgu.ac.il, eliyanac@bgu.ac.il}
}

\begin{document}

\maketitle

\begin{abstract}
Neural audio codecs compress waveforms into compact discrete tokens that underpin speech language models, real-time communication, and large-scale audio storage. Almost every dominant design, including residual vector quantization, finite scalar quantization, and single-codebook variants, follows the VQ-VAE template by partitioning the encoder latent through a learned codebook or a fixed scalar grid. We ask whether this partition is necessary. We introduce GS-Codec, a neural speech codec whose bottleneck is a parametric signal decomposition rather than a quantizer. We adapt Gaussian splatting from 3D scene reconstruction to one-dimensional latents. An inner optimization loop fits each encoder segment as a weighted sum of 1D Gaussian primitives. The decoder then reconstructs the waveform from the rendered sum. To avoid the cost of this iterative inner loop at inference time, we additionally train a lightweight GS Predictor Net that regresses the primitive parameters in a single forward pass. The encoder and decoder are trained end-to-end through the inner loop, with no quantizer anywhere in the training pipeline: the bottleneck is the decomposition itself, and scalar quantization is applied only post-training to the fitted parameters. Rather than relying on discrete codebook stages for bitrate control, our representation exposes a fine-grained rate-quality tradeoff: a single trained checkpoint supports post-training bitrate control by varying the number of primitives and the per-parameter bit depth, with no retraining required. GS-Codec matches or exceeds well-established open-source codecs such as EnCodec and DAC on speaker similarity (SIM), intelligibility (STOI), and perceptual quality (UTMOS) at comparable bitrates, while achieving comparable semantic performance (WER). Code and audio samples are available at \url{https://ronaluf.github.io/gs-codec/}.
\end{abstract}

\section{Introduction}
\label{sec:intro}

In the wake of Large Language Models~\citep{brown2020language,transformer,bert,touvron2023llama,touvron2023llama2,openai2023gpt4},
there is growing demand for unified multimodal architectures and efficient audio compression.
Neural audio codecs have emerged as foundational technology, compressing raw waveforms via neural networks~\citep{soundstream,encodec,dac} and tokenizing them into compact discrete representations that bridge high-rate acoustic signals and transformer-based models~\citep{audiolm,speechtokenizer}.
These tokens are essential for TTS~\citep{valle,valle2,voicebox,soundstorm,kharitonov2023speak,shen2024naturalspeech2,naturalspeech3},
music generation~\citep{musiclm,musicgen,audiolm,audiogen,discoder},
voice conversion, dialogue~\citep{moshi,speechtokenizer,wavtokenizer,cosyvoice},
and real-time communication~\citep{soundstream,lyrav2,opus}.
Token fidelity directly bounds synthesis quality in LM-based TTS~\citep{valle,audiolm} and music generation~\citep{musiclm,musicgen}, making codecs the critical infrastructure for multimodal audio systems~\citep{moshi,zhan2024anygpt,audiopalm}.

The dominant architecture~\citep{encodec,dac,hificodec,snac} pairs a convolutional encoder with a Residual Vector Quantization (RVQ) bottleneck~\citep{soundstream} and mirror decoder.
RVQ iteratively quantizes residual error across codebook stages, enabling flexible bitrate-quality trade-offs; EnCodec~\citep{encodec} and DAC~\citep{dac} achieve near-transparent reconstruction at 3-6\,kbps.
However, these models share a structural assumption: compression must proceed through a learned discrete codebook, committing the latent to a lookup over a finite set of vectors fixed at training time.
Bitrate adjusts only in discrete steps defined by the number of codebook stages, and the representation is susceptible to codebook collapse~\citep{jukebox,robustvq,simvq}, wasting representational capacity.
Whether the bottleneck can instead be a continuous parametric decomposition that encodes structure rather than assigns indices remains an open question.

We introduce \textbf{GS-Codec}, a neural audio codec that replaces the RVQ quantizer with a Gaussian Splatting (GS) bottleneck.
Building on Gaussian splatting's success in image representation and compression~\citep{gaussianimage,gaussianimagepp,imagegs,instantgaussianimage}, GS-Codec fits $N_G$ localized Gaussian primitives to each audio segment's latent via a lightweight per-clip optimizer.
Each primitive shares a temporal width and per-channel amplitudes across a local neighbourhood of frames. This overlapping representation provides a structural prior at the bottleneck that framewise quantization does not explicitly impose.
Rather than mapping to discrete codes (RVQ) or a fixed scalar grid (FSQ), GS-Codec uses a parametric signal decomposition: the encoder latent is approximated as a weighted sum of Gaussian primitives, with only those parameters scalar-quantized post-training.
Since quantization is applied after training, a single checkpoint supports fine-grained post-training bitrate control by varying the number of primitives and per-parameter bit depth, with no retraining required.
Experiments on LibriTTS, LJSpeech, and LibriSpeech show GS-Codec matches or exceeds established open-source codecs on perceptual quality and speaker similarity, while opening new design axes for neural audio compression.

Our contributions are as follows:
\begin{enumerate}
\setlength{\itemsep}{2pt}\setlength{\parsep}{0pt}
\item We propose GS-Codec, a neural speech codec whose bottleneck is a 1D Gaussian-splatting decomposition rather than a vector or scalar quantizer. To our knowledge, this is the first neural codec to replace partition-based discretization with a parametric signal decomposition inside the latent bottleneck.

\item We introduce a lightweight GS Predictor Net that learns to approximate the iterative Gaussian Splatting process, amortizing the multi-step optimization loop into a single forward pass for fast inference.
\item We demonstrate fine-grained post-training bitrate control via varying primitive count and bit depth, and report comparisons against open-source speech codecs on LibriTTS, LJSpeech, and LibriSpeech, matching or exceeding baselines on perceptual quality and speaker similarity.
\end{enumerate}

\section{Related Work}
\label{sec:related}

Neural speech and audio coding follows an encoder-quantiser-decoder template.
WaveNet~\citep{wavenet} motivated faster alternatives displacing Opus~\citep{opus} and LPCNet~\citep{lpcnet};
SoundStream~\citep{soundstream}, Lyra~V2~\citep{lyrav2}, EnCodec~\citep{encodec}, and DAC~\citep{dac} established RVQ as the dominant bottleneck.
Adversarial and diffusion vocoders-HiFi-GAN~\citep{hifigan}, MelGAN~\citep{melgan}, Parallel WaveGAN~\citep{parallelwavegan}, UnivNet~\citep{univnet}, BigVGAN~\citep{bigvgan}, Vocos~\citep{vocos}, DiffWave~\citep{diffwave}-showed that multi-scale spectrogram losses drive fidelity gains later adopted by codec pipelines~\citep{wavtokenizer,hificodec}.
Variants target quantiser structure (HiFi-Codec~\citep{hificodec}, SuperCodec~\citep{supercodec}, SNAC~\citep{snac}), benchmarking~\citep{codecsuperb}, semantic alignment~\citep{speechtokenizer}, single-codebook compression~\citep{wavtokenizer}, real-time dialogue~\citep{moshi}, and dynamic frame rates~\citep{flexicodec}.
Ultra-low-rate work includes DualCodec~\citep{dualcodec} (12.5\,Hz dual-stream), SemantiCodec~\citep{semanticodec} (diffusion decoder, sub-kbps), TAAE~\citep{taae} (FSQ, 400-700\,bps), TaDiCodec~\citep{tadicodec} (6.25\,Hz, text-guided diffusion), TS3-Codec~\citep{ts3codec} (pure transformer), X-Codec~\citep{xcodec2}, and NaturalSpeech\,3~\citep{naturalspeech3} (factorised VQ).

The VQ bottleneck traces to~\citet{gray_vq} and VQ-VAE~\citep{vqvae,vqvae2}, with extensions for collapse~\citep{jukebox}, robustness~\citep{robustvq}, residual composition~\citep{rqvae,improvedrvq}, inter-codebook optimisation~\citep{ervq}, regularisation~\citep{simvq}, variable bitrates~\citep{variablervq}, and FSQ~\citep{fsq}, which quantizes each latent coordinate on a fixed scalar grid instead of a learned codebook-all via straight-through gradients~\citep{ste}; our bottleneck requires no codebook and no straight-through estimator for discretization, as scalar quantization is applied only post-training to the continuous fitted parameters~\citep{lloyd1982}.

Codec tokens underpin language models~\citep{audiolm,valle,valle2,musiclm,musicgen,audiogen,soundstorm,voicebox,cosyvoice,audiopalm} built on transformers~\citep{transformer}, self-supervised encoders~\citep{cpc,wav2vec2,hubert,w2vbert,wavlm,whisper,data2vec}, and multimodal architectures~\citep{zhan2024anygpt}; token quality directly bounds downstream fidelity~\citep{gslm}.
Cross-codec benchmarks~\citep{codecsuperb,superb,bert} and metrics PESQ~\citep{pesq}, UTMOS~\citep{utmos}, STOI~\citep{stoi} anchor evaluation.

Our bottleneck is a parametric signal decomposition rather than a quantiser, following the differentiable DSP tradition~\citep{ddsp,ddsp_review} but applied inside the bottleneck.
Gaussian splatting~\citep{3dgs,nerf} has been extended to 2-D geometry~\citep{gs2d}, image compression~\citep{gaussianimage,gaussianimagepp,imagegs,instantgaussianimage,llgaussianimage}, and 1-D time-series~\citep{timegs}; broader radial-basis kernels~\citep{darbsplat} suggest natural extensions.
Feed-forward amortisation~\citep{splatterimage,pixelsplat,latentsplat}, implicit compression~\citep{coinpp,functa,learnedinit}, and Instant-NGP~\citep{instantngp} inform our inner-loop design.
SIRENs~\citep{siren} ground per-signal fitting; \citet{siamese_siren} apply this to audio by fitting a SIREN per clip-we share that spirit but replace the network with an analytic Gaussian decomposition, making parameters directly quantisable.
Rate control contemporaries FlexiCodec~\citep{flexicodec} and WavTokenizer~\citep{wavtokenizer} rely on fixed codebooks; GS-Codec instead varies the resolution of a parametric splat fit.
Our SEANet~\citep{tagliasacchi2020seanet} backbone uses the Snake activation~\citep{snake} and AdamW~\citep{adamw}, with a VAE/GAN~\citep{vae,gan} training skeleton.

\section{Method}
\label{sec:method}

\begin{figure}[h]
\centering
\includegraphics[width=\linewidth]{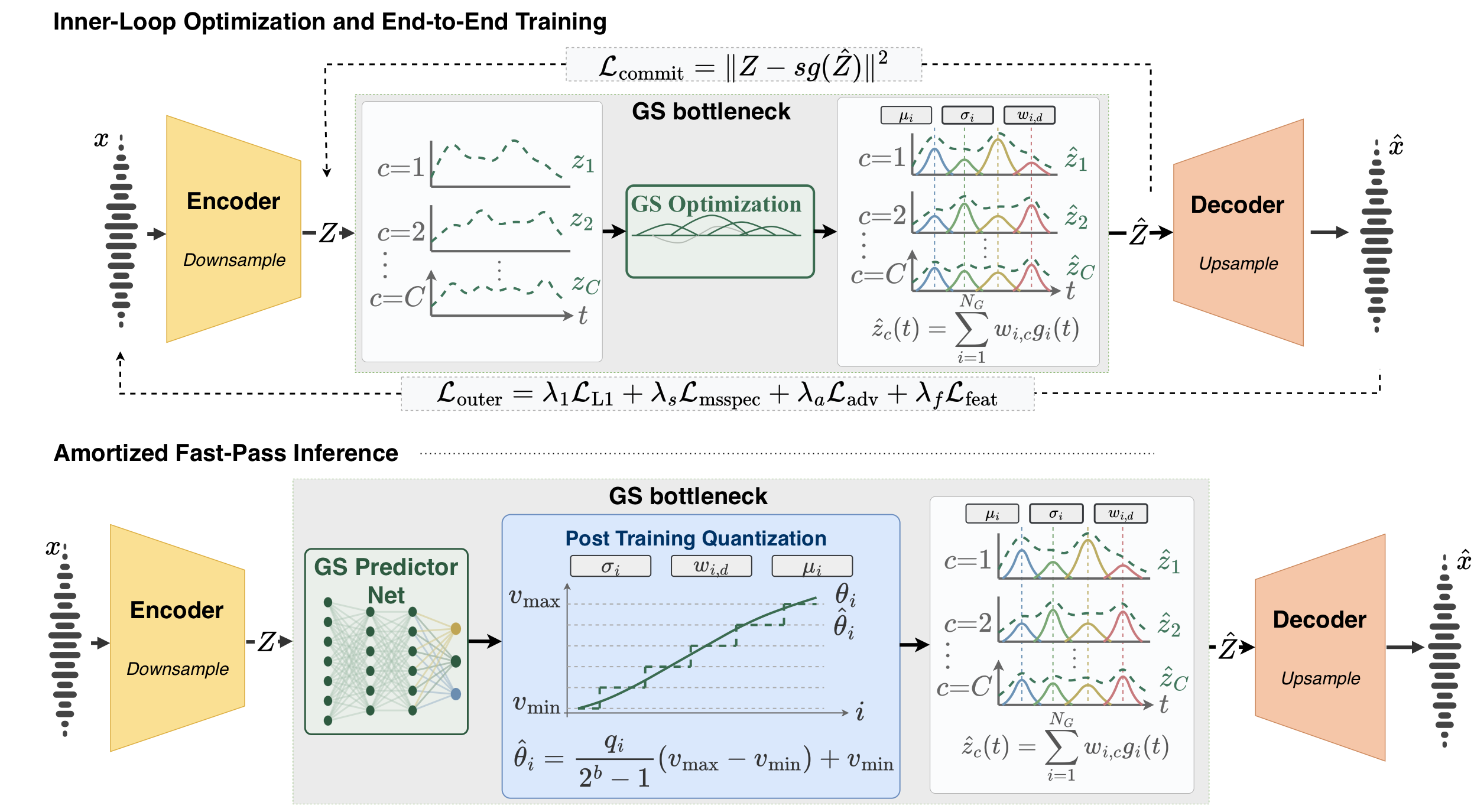}
\caption{GS-Codec pipeline. \textbf{Training} (top): The encoder latent $Z$ is approximated by $N_G$ Gaussian primitives via an inner optimization loop. \textbf{Inference} (bottom): A GS Predictor Net regresses primitive parameters in a single pass, which are then quantized, transmitted, and decoded.}
\label{fig:arch}
\end{figure}

We propose GS-Codec, a neural audio compression method whose bottleneck is a parametric signal decomposition based on one-dimensional Gaussian splatting. An encoder network, denoted as $E$ and based on the SEANet architecture \citep{soundstream}, maps the input waveform to a continuous latent representation. An inner optimization loop then fits this latent as a weighted sum of one-dimensional Gaussian primitives. A SEANet-based decoder network, denoted as $D$, then reconstructs the audio from this parametric approximation. The entire model is trained end-to-end, with $E$ learning to yield splattable representations and $D$ learning to invert the rendered sum. The overall architecture is depicted in Figure~\ref{fig:arch}.

\subsection{End-to-End Training with the GS Bottleneck}
\label{sec:method:training}

The training pipeline consists of the encoder $E$, the Gaussian-splatting bottleneck, and the decoder $D$, optimized jointly via a combination of reconstruction and adversarial losses.

\subsubsection{Architecture}
\label{sec:method:architecture}

\paragraph{Encoder}
The encoder network $E$ maps a time-domain input waveform $x \in \mathbb{R}^{1 \times L}$, where $L$ is the number of audio samples, to a continuous latent tensor $Z \in \mathbb{R}^{C \times T}$, where $C$ is the channel dimension and $T$ is the number of temporal latent frames. The network $E$ follows the SEANet convolutional architecture \citep{soundstream} with the Snake activation \citep{snake}, progressively downsampling the input waveform through strided convolutions to a lower-rate latent representation. Concrete values for $C$ and other architectural hyperparameters are given in Appendix~\ref{app:details}.

\paragraph{GS Bottleneck and 1D Gaussian Splatting}
The core of GS-Codec is the Gaussian-splatting bottleneck. Inspired by 3D Gaussian splatting for scene reconstruction \citep{3dgs}, we adapt the formulation to one-dimensional latent fitting. To the best of our knowledge, this is the first application of Gaussian splatting as a compression bottleneck in a neural audio codec. A Gaussian primitive is a spatially localized, one-dimensional basis function parameterized by a temporal center $\mu \in \mathbb{R}$ and a temporal scale $\sigma \in \mathbb{R}^+$. The primitive is defined as $\phi(t; \mu, \sigma) = \exp\left(-\frac{(t - \mu)^2}{2\sigma^2}\right)$.
The basis function $\phi$ is differentiable everywhere and bounded, making it well-suited for gradient-based optimization.

To represent a segment of the latent signal $Z$, we approximate it as a weighted superposition of $N_G$ Gaussian primitives, where $N_G$ is the number of primitives. Each primitive $i \in \{1, \dots, N_G\}$ is described by its center $\mu_i$, its scale $\sigma_i$ (which is shared across all $C$ channels), and a per-channel amplitude vector $w_i \in \mathbb{R}^C$. Thus, the rendered approximation for a specific channel $c \in \{1, \dots, C\}$ at time frame $t \in \{0, \dots, T{-}1\}$ is computed as $\hat{Z}_{c,t} = \sum_{i=1}^{N_G} w_{i,c} \phi(t; \mu_i, \sigma_i)$,
which we denote collectively as the rendered tensor $\hat{Z} \in \mathbb{R}^{C \times T}$.

\begin{figure}[h]
\centering
\includegraphics[width=0.6\linewidth]{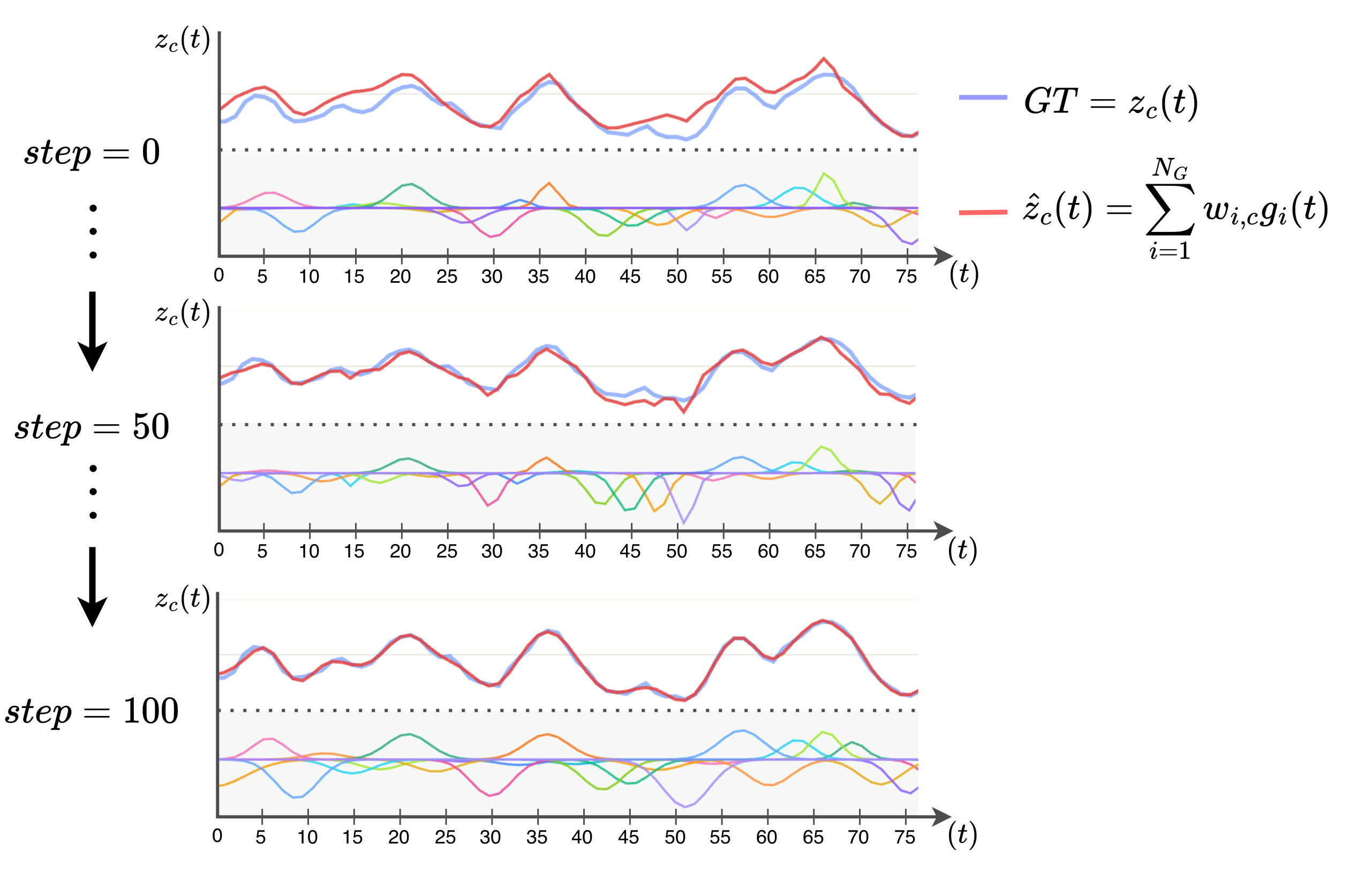}
\caption{Visualization of the GS inner loop on a single latent channel across optimization steps 0, 50, and 100. Ground-truth $Z_c$ is shown in blue, rendered approximation $\hat{Z}_c$ in red, and individual weighted Gaussian primitives are shown underneath. The horizontal axis denotes the temporal latent frame ($t$), while the vertical axis represents the latent value.}
\label{fig:gs_detail}
\end{figure}

The parameters of these primitives, denoted as $\Theta = \{\mu_i, \sigma_i, w_i\}_{i=1}^{N_G}$, are obtained via a 1D Gaussian splatting inner loop applied during the forward pass. Given a fixed latent $Z$ from the encoder $E$, we minimize the mean squared error (MSE) loss, defined as
\begin{equation}
\label{eq:loss_inner}
\mathcal{L}_{\mathrm{inner}} = \frac{1}{CT}\|Z - \hat{Z}\|_F^2.
\end{equation}
Starting from an initial state $\Theta^{(0)}$, the parameters are iteratively updated for $K$ gradient-based optimization steps. While any first-order optimizer (e.g., Adam \citep{adam}) can be used in practice, a standard gradient descent step takes the form
\begin{equation}
\label{eq:update}
\Theta^{(k)} = \Theta^{(k-1)} - \alpha^{(k)} \nabla_{\Theta} \mathcal{L}_{\mathrm{inner}}(\Theta^{(k-1)}), \quad k \in \{1, \dots, K\},
\end{equation}
where $\alpha^{(k)}$ is the step size. Figure~\ref{fig:gs_detail} illustrates this convergence on a single latent channel.

\paragraph{Decoder}
The decoder network $D$ mirrors the encoder architecture, applying transposed convolutions to progressively upsample the reconstructed latent tensor $\hat{Z} \in \mathbb{R}^{C \times T}$ back to the original sample rate. The output is a reconstructed waveform $\hat{x} \in \mathbb{R}^{1 \times L}$, where $L$ matches the input length. The decoder uses the same channel widths, number of residual blocks, and Snake activation \citep{snake} as the encoder, following the SEANet design.

\subsubsection{Loss Functions}
\label{sec:method:losses}

The model is trained using a combination of objectives applied to both the intermediate latent space and the final synthesized waveform.
To stabilize training and ensure the encoder latents remain suitable for splatting, we apply a commitment loss inside the bottleneck, defined as $\mathcal{L}_{\mathrm{commit}} = \| Z - \mathrm{sg}(\hat{Z}) \|_2^2$,
where $\mathrm{sg}(\cdot)$ denotes the stop-gradient operation. This pulls the encoder output toward the splatted reconstruction it has to match.
Outer gradients from $\mathcal{L}$ reach the encoder $E$ via a straight-through estimator: the decoder receives $\hat{Z}_{\mathrm{STE}} = Z + \mathrm{sg}(\hat{Z} - Z)$, which has the value of the splat $\hat{Z}$ but carries the gradient of the encoder output $Z$.

The synthesized waveform $\hat{x}$ is supervised using an outer reconstruction loss comprising an L1 time-domain term $\mathcal{L}_{\mathrm{L1}}$, a multi-scale spectrogram term $\mathcal{L}_{\mathrm{msspec}}$, an adversarial loss $\mathcal{L}_{\mathrm{adv}}$, and a feature-matching loss $\mathcal{L}_{\mathrm{feat}}$ from a multi-scale Short-Time Fourier Transform (STFT) discriminator. The total training loss combines all objectives:
\begin{equation}
\label{eq:loss_outer}
\mathcal{L} = \lambda_1 \mathcal{L}_{\mathrm{L1}} + \lambda_s \mathcal{L}_{\mathrm{msspec}} + \lambda_a \mathcal{L}_{\mathrm{adv}} + \lambda_f \mathcal{L}_{\mathrm{feat}} + \lambda_c \mathcal{L}_{\mathrm{commit}},
\end{equation}
where $\lambda_1, \lambda_s, \lambda_a, \lambda_f, \lambda_c \in \mathbb{R}^+$ are scalar weighting hyperparameters (concrete values in Appendix~\ref{app:details}). The gradients from $\mathcal{L}$ are backpropagated through $D$ and $E$ via a gradient balancer.

\subsection{Inference}
\label{sec:method:inference}

During inference, the model compresses audio segments into a set of quantized parameters, which are then transmitted and decoded to reconstruct the waveform.

\subsubsection{Parameter Quantization and Token Format}
\label{sec:method:quantization}

To achieve compression, each stored parameter of the fitted Gaussians is scalar-quantized to $B$ bits for transmission, where $B$ is the bit depth. The quantization levels are determined via $k$-means clustering on a held-out calibration subset after training is complete: for each parameter type (scales $\sigma_i$ and amplitudes $w_{i,c}$ separately), we fit $2^B$ centroids to the empirical distribution of full-precision values collected from the calibration set. At inference time, each parameter is mapped to its nearest centroid index for transmission, and the receiver looks up the centroid value to dequantize.

Positions are handled in one of two ways. In the frozen-position variant, the centers lie on a uniform temporal grid, $\mu_i = (i-1)(T-1)/(N_G-1)$ for $i\in\{1,\ldots,N_G\}$ and $N_G>1$; each $\mu_i$ is a fixed scalar that does not vary with the temporal index $t$, and no positions are transmitted. In the learned-position variant, positions are optimized per segment and transmitted as additional parameters. Section~\ref{sec:ablations} compares the two. With frozen positions, the transmitted parameters per segment consist of $N_G$ scale values $\sigma_i$ and $N_G \times C$ amplitude values $w_{i,c}$, each quantized to $B$ bits, giving a theoretical bitrate of
\begin{equation}
\label{eq:bitrate}
\mathrm{BR}_{\mathrm{kbps}} = \frac{(C + 1) N_G B}{T_{\mathrm{seg}} \times 1000},
\end{equation}
where $T_{\mathrm{seg}}$ is the duration of the audio segment in seconds. Both $N_G$ and $B$ are integers, so the supported rates are discrete. When positions are learned and transmitted, each $\mu_i$ is quantized independently to $B_\mu$ bits (a separate budget from $B$), and the bitrate becomes $\bigl((C+1)N_G B + N_G B_\mu\bigr) / (T_{\mathrm{seg}} \times 1000)$.

\subsubsection{GS Predictor Net}
\label{sec:method:gspredictor}

The inner optimization loop described above requires $K$ gradient steps per segment at encoding time, which is substantially slower than a feed-forward codebook lookup. To address this, we train a GS Predictor Net $P$ that amortises the inner loop, regressing the Gaussian parameters from the continuous latent $Z$ in a single forward pass.

\paragraph{Architecture.}
The predictor takes the encoder latent $Z \in \mathbb{R}^{C \times T}$ as input and outputs the full parameter set: $N_G$ scale values $\sigma_i$, the $N_G \times C$ amplitude matrix $w_{i,c}$, and $N_G$ position values $\mu_i$. A stack of dilated 1D convolutional blocks first extracts temporal features from $Z$. A transformer decoder then cross-attends from $N_G$ learned queries, one per primitive into these features, producing a per-primitive representation. Three linear output heads map each query to its scale (Softplus), amplitude vector, and position (Sigmoid scaled to $[0, T{-}1]$). The predictor runs with predicted or frozen centers; in the frozen case the position head is bypassed. Its queries can also be generated from a shared learned token and positional encodings of an $N_G$-point grid, so one checkpoint serves different $N_G$.

\paragraph{Training.}
The predictor is trained after the main encoder-decoder training is complete, with the encoder and decoder weights frozen. The objective is a simple MSE loss between the rendered splat produced from the predicted parameters and the encoder latent:
\begin{equation}
\label{eq:loss_predictor}
\mathcal{L}_{\mathrm{pred}} = \| \hat{Z}_{\mathrm{splat}} - Z \|_2^2,
\end{equation}
where $\hat{Z}_{\mathrm{splat}}$ is obtained by evaluating the rendered approximation with the predicted parameters. The position head uses a higher learning rate than the remaining parameters to accelerate convergence of the temporal placement.

\paragraph{Inference.}
At inference time, the predictor replaces the inner loop entirely: the encoder produces $Z$, the predictor outputs the primitive parameters in a single pass, and these are scalar-quantized and transmitted. The decoder side is unchanged. This amortisation reduces the encoding cost from $K$ optimization steps to a single forward pass through the predictor, while the reconstructed waveform quality remains close to that of the full iterative fit.

\section{Experiments}
\label{sec:exp}

\subsection{Experimental setup}
\label{sec:setup}

\paragraph{Datasets.}
For the main comparison with WavTokenizer-protocol baselines (Table~\ref{tab:wavtokenizer_comparison}), we train GS-Codec on Common Voice~\citep{commonvoice} (${\approx}1{,}717$\,h), LibriTTS~\citep{libritts} (${\approx}554$\,h), and VCTK~\citep{vctk} (${\approx}68$\,h), all resampled to 24\,kHz monaural (${\approx}2{,}339$\,h total).
Speech reconstruction is evaluated on LibriTTS \texttt{test-clean} (clean conditions), LibriTTS \texttt{test-other} (noisy conditions), and LJSpeech (out-of-domain generalisation).
For Table~\ref{tab:comparison_baselines}, we train on the 54{,}000-hour Librilight-Large dataset~\citep{kahn2020librilight}, resampled to 16\,kHz.
We evaluate on LibriSpeech \texttt{test-clean}~\citep{librispeech}.
For the ablation studies (\S\ref{sec:sweep}), all models are trained on the same 2{,}339-hour mixture and evaluated on LibriSpeech \texttt{test-clean}.

\paragraph{Baselines.}
For Table~\ref{tab:wavtokenizer_comparison} we compare against EnCodec~\citep{encodec}, DAC~\citep{dac}, SpeechTokenizer~\citep{speechtokenizer}, HiFi-Codec~\citep{hificodec}, Vocos~\citep{vocos}, and WavTokenizer~\citep{wavtokenizer}, following the evaluation protocol of \citet{wavtokenizer}. Baseline numbers are quoted directly from \citet{wavtokenizer}.
For Table~\ref{tab:comparison_baselines} we additionally compare against Mimi~\citep{moshi} and XYTokenizer~\citep{gong2025xytokenizer}. 
\paragraph{Metrics.}
We report UTMOS~\citep{utmos} as the primary perceptual quality metric, alongside PESQ~\citep{pesq}, STOI~\citep{stoi}, and V/UV F1.
Wideband PESQ\,(WB) is used for the WavTokenizer comparison (Table~\ref{tab:wavtokenizer_comparison}); narrowband PESQ\,(NB) for Table~\ref{tab:comparison_baselines}.
For Table~\ref{tab:comparison_baselines} we additionally report VALL-E speaker similarity (SIM)~\citep{valle} and WER, computed from transcriptions by \texttt{facebook/hubert-large-ls960-ft}~\citep{hubert} on the 4-10\,s subset ($n{=}1{,}237$).

\paragraph{Hyperparameters.}
The SEANet encoder/decoder uses $C{=}32$ latent channels, 64 base filters,
and Snake activations~\citep{snake}, producing 225 latent frames per
3-second segment at 75\,Hz. We train with Adam ($\beta_1{=}0.5$,
$\beta_2{=}0.9$, LR $10^{-4}$) and EMA (decay $0.99$) for 200 epochs of
2{,}000 steps on a single NVIDIA RTX PRO 6000 GPU (${\approx}144$\,h).
The GS bottleneck uses $N_G{=}80$ primitives per 3-second segment and $K{=}100$ inner Adam
steps during training. At evaluation time we run 300 steps under a cosine
schedule. Positions are fixed to a uniform grid; parameters are
scalar-quantised with per-type $k$-means codebooks ($2^B$ centroids)
calibrated on \texttt{libritts-dev-clean}.
We report three decode-time variants for each $(N_G, B)$ configuration: \textbf{unfrozen} - positions are optimised per-utterance and transmitted at a fixed $B_\mu{=}10$ bits, yielding a slight bitrate increase; \textbf{frozen} - positions are fixed to a uniform grid and not transmitted (default); and \textbf{Pred} - the GS Predictor Network (\S\ref{sec:method:gspredictor}) replaces iterative optimisation at encoding time.
The two iterative variants serve as upper bounds; \textbf{Pred} is the deployable single-pass codec.

\subsection{Results}
\label{sec:results}

\begin{table}[p]
  \caption{Objective reconstruction results on LibriTTS \texttt{test-clean} (clean),
  LibriTTS \texttt{test-other} (noisy), and LJSpeech (out-of-domain), following the
  protocol of \citet{wavtokenizer} Table~1.
  PESQ is wideband (WB). Best results in each band are \textbf{bold}.}
  \label{tab:wavtokenizer_comparison}
  \centering
  \scriptsize
  \renewcommand{\arraystretch}{1.0}
  \setlength{\tabcolsep}{3pt}
  \begin{tabularx}{\textwidth}{ll X cccc cc}
    \toprule
    \textbf{Dataset} & \textbf{Band} & \textbf{Model} & \textbf{BW\,$\downarrow$} & \textbf{tok/s\,$\downarrow$}
    & \textbf{UTMOS\,$\uparrow$} & \textbf{PESQ\,$\uparrow$} & \textbf{STOI\,$\uparrow$} & \textbf{V/UV F1\,$\uparrow$} \\
    \midrule
    \multirow{22}{*}{\hspace{-0.4cm}\rotatebox{90}{\textit{LibriTTS test-clean}}}
    & & DAC (ref) & 9.0 kbps & 900 & 3.910 & 3.908 & 0.970 & 0.978 \\
    \cmidrule(l){2-9}
    & \multirow{6}{*}{\rotatebox{90}{\begin{tabular}{@{}c@{}}\textbf{$\geq$5\,kbps} \\ high\end{tabular}}} & EnCodec & 6.0 kbps & 600 & 3.040 & 2.720 & 0.939 & 0.953 \\
    & & Vocos & 6.0 kbps & 600 & 3.695 & 2.807 & 0.943 & 0.944 \\
    & & SpeechTokenizer & 6.0 kbps & 600 & 3.879 & 2.612 & 0.917 & 0.950 \\
    & & \gc{GS-Codec ($N_G{=}90$, 6b, Pred, frozen)} & \gc{5.94 kbps} & \gc{990} & \gc{4.076} & \gc{3.289} & \gc{0.970} & \gc{0.956} \\
    & & \gc{GS-Codec ($N_G{=}90$, 6b, frozen)} & \gc{5.94 kbps} & \gc{990} & \gc{4.094} & \gc{3.348} & \gc{0.972} & \gc{\textbf{0.960}} \\
    & & \gc{\textbf{GS-Codec ($N_G{=}90$, 6b, unfrozen)}} & \gc{6.24 kbps} & \gc{1020} & \gc{\textbf{4.102}} & \gc{\textbf{3.400}} & \gc{\textbf{0.974}} & \gc{0.958} \\
    \cmidrule(l){2-9}
    & \multirow{9}{*}{\rotatebox{90}{\begin{tabular}{@{}c@{}}\textbf{3-5\,kbps} \\ mid\end{tabular}}} & DAC & 4.0 kbps & 400 & 3.433 & 2.738 & 0.928 & 0.957 \\
    & & HiFi-Codec & 3.0 kbps & 400 & 3.753 & 2.961 & 0.941 & \textbf{0.962} \\
    & & HiFi-Codec & 4.0 kbps & 300 & 3.904 & 3.012 & 0.945 & 0.958 \\
    & & EnCodec & 3.0 kbps & 300 & 2.307 & 2.052 & 0.901 & 0.920 \\
    & & Vocos & 3.0 kbps & 300 & 3.539 & 2.403 & 0.923 & 0.936 \\
    & & SpeechTokenizer & 3.0 kbps & 300 & 3.563 & 1.931 & 0.878 & 0.927 \\
    & & \gc{GS-Codec ($N_G{=}80$, 4b, Pred, frozen)} & \gc{3.52 kbps} & \gc{880} & \gc{3.990} & \gc{3.071} & \gc{0.959} & \gc{0.943} \\
    & & \gc{GS-Codec ($N_G{=}80$, 4b, frozen)} & \gc{3.52 kbps} & \gc{880} & \gc{3.974} & \gc{3.067} & \gc{0.960} & \gc{0.950} \\
    & & \gc{\textbf{GS-Codec ($N_G{=}80$, 4b, unfrozen)}} & \gc{3.79 kbps} & \gc{907} & \gc{\textbf{4.026}} & \gc{\textbf{3.177}} & \gc{\textbf{0.966}} & \gc{0.953} \\
    \cmidrule(l){2-9}
    & \multirow{6}{*}{\rotatebox{90}{\begin{tabular}{@{}c@{}}\textbf{${<}$3\,kbps} \\ low\end{tabular}}} & DAC & 1.0 kbps & 100 & 1.494 & 1.246 & 0.771 & 0.794 \\
    & & WavTokenizer & 0.5 kbps & 40 & 3.602 & 1.703 & 0.862 & 0.917 \\
    & & WavTokenizer & 0.9 kbps & 75 & \textbf{4.049} & 2.373 & 0.914 & 0.938 \\
    & & \gc{\textbf{GS-Codec ($N_G{=}50$, 4b, unfrozen)}} & \gc{2.37 kbps} & \gc{567} & \gc{3.521} & \gc{2.409} & \gc{0.932} & \gc{0.930} \\
    & & \gc{GS-Codec ($N_G{=}60$, 4b, frozen)} & \gc{2.64 kbps} & \gc{660} & \gc{3.568} & \gc{2.453} & \gc{0.929} & \gc{0.933} \\
    & & \gc{\textbf{GS-Codec ($N_G{=}60$, 4b, unfrozen)}} & \gc{2.84 kbps} & \gc{680} & \gc{3.847} & \gc{\textbf{2.825}} & \gc{\textbf{0.951}} & \gc{\textbf{0.940}} \\
    \midrule
    \multirow{22}{*}{\hspace{-0.4cm}\rotatebox{90}{\textit{LibriTTS test-other}}}
    & & DAC (ref) & 9.0 kbps & 900 & 3.357 & 3.760 & 0.958 & 0.970 \\
    \cmidrule(l){2-9}
    & \multirow{6}{*}{\rotatebox{90}{\begin{tabular}{@{}c@{}}\textbf{$\geq$5\,kbps} \\ high\end{tabular}}} & EnCodec & 6.0 kbps & 600 & 2.657 & 2.682 & 0.924 & 0.934 \\
    & & Vocos & 6.0 kbps & 600 & 3.196 & 2.559 & 0.921 & 0.920 \\
    & & SpeechTokenizer & 6.0 kbps & 600 & 3.285 & 2.327 & 0.881 & 0.921 \\
    & & \gc{GS-Codec ($N_G{=}90$, 6b, Pred, frozen)} & \gc{5.94 kbps} & \gc{990} & \gc{3.623} & \gc{3.011} & \gc{0.955} & \gc{0.939} \\
    & & \gc{GS-Codec ($N_G{=}90$, 6b, frozen)} & \gc{5.94 kbps} & \gc{990} & \gc{3.631} & \gc{3.080} & \gc{0.958} & \gc{0.943} \\
    & & \gc{\textbf{GS-Codec ($N_G{=}90$, 6b, unfrozen)}} & \gc{6.24 kbps} & \gc{1020} & \gc{\textbf{3.639}} & \gc{\textbf{3.131}} & \gc{\textbf{0.960}} & \gc{\textbf{0.943}} \\
    \cmidrule(l){2-9}
    & \multirow{9}{*}{\rotatebox{90}{\begin{tabular}{@{}c@{}}\textbf{3-5\,kbps} \\ mid\end{tabular}}} & DAC & 4.0 kbps & 400 & 2.945 & 2.595 & 0.908 & 0.940 \\
    & & HiFi-Codec & 4.0 kbps & 400 & 3.075 & 2.554 & 0.913 & 0.939 \\
    & & HiFi-Codec & 3.0 kbps & 300 & 3.303 & 2.608 & 0.917 & 0.932 \\
    & & EnCodec & 3.0 kbps & 300 & 2.088 & 2.052 & 0.884 & 0.893 \\
    & & Vocos & 3.0 kbps & 300 & 3.056 & 2.193 & 0.897 & 0.905 \\
    & & SpeechTokenizer & 3.0 kbps & 300 & 3.018 & 1.737 & 0.837 & 0.891 \\
    & & \gc{GS-Codec ($N_G{=}80$, 4b, Pred, frozen)} & \gc{3.52 kbps} & \gc{880} & \gc{3.455} & \gc{2.786} & \gc{0.938} & \gc{0.932} \\
    & & \gc{GS-Codec ($N_G{=}80$, 4b, frozen)} & \gc{3.52 kbps} & \gc{880} & \gc{3.504} & \gc{2.831} & \gc{0.944} & \gc{0.934} \\
    & & \gc{\textbf{GS-Codec ($N_G{=}80$, 4b, unfrozen)}} & \gc{3.79 kbps} & \gc{907} & \gc{\textbf{3.539}} & \gc{\textbf{2.907}} & \gc{\textbf{0.950}} & \gc{\textbf{0.942}} \\
    \cmidrule(l){2-9}
    & \multirow{6}{*}{\rotatebox{90}{\begin{tabular}{@{}c@{}}\textbf{${<}$3\,kbps} \\ low\end{tabular}}} & DAC & 1.0 kbps & 100 & 1.499 & 1.245 & 0.751 & 0.778 \\
    & & WavTokenizer & 0.5 kbps & 40 & 3.055 & 1.662 & 0.834 & 0.895 \\
    & & WavTokenizer & 0.9 kbps & 75 & \textbf{3.431} & 2.261 & 0.891 & 0.917 \\
    & & \gc{\textbf{GS-Codec ($N_G{=}50$, 4b, unfrozen)}} & \gc{2.37 kbps} & \gc{567} & \gc{3.117} & \gc{2.310} & \gc{0.915} & \gc{0.926} \\
    & & \gc{GS-Codec ($N_G{=}60$, 4b, frozen)} & \gc{2.64 kbps} & \gc{660} & \gc{3.130} & \gc{2.331} & \gc{0.912} & \gc{0.926} \\
    & & \gc{\textbf{GS-Codec ($N_G{=}60$, 4b, unfrozen)}} & \gc{2.84 kbps} & \gc{680} & \gc{3.378} & \gc{\textbf{2.620}} & \gc{\textbf{0.934}} & \gc{\textbf{0.934}} \\
    \midrule
    \multirow{20}{*}{\hspace{-0.4cm}\rotatebox{90}{\textit{LJSpeech}}}
    & & DAC (ref) & 9.0 kbps & 900 & 4.301 & 3.902 & 0.973 & 0.976 \\
    \cmidrule(l){2-9}
    & \multirow{6}{*}{\rotatebox{90}{\begin{tabular}{@{}c@{}}\textbf{$\geq$5\,kbps} \\ high\end{tabular}}} & EnCodec & 6.0 kbps & 600 & 3.229 & 2.663 & 0.944 & \textbf{0.956} \\
    & & Vocos & 6.0 kbps & 600 & 4.033 & 2.926 & 0.950 & 0.946 \\
    & & SpeechTokenizer & 6.0 kbps & 600 & 4.237 & 2.641 & 0.932 & 0.945 \\
    & & \gc{GS-Codec ($N_G{=}90$, 6b, Pred, frozen)} & \gc{5.94 kbps} & \gc{990} & \gc{4.346} & \gc{3.205} & \gc{0.970} & \gc{0.952} \\
    & & \gc{GS-Codec ($N_G{=}90$, 6b, frozen)} & \gc{5.94 kbps} & \gc{990} & \gc{4.347} & \gc{3.273} & \gc{0.971} & \gc{0.944} \\
    & & \gc{\textbf{GS-Codec ($N_G{=}90$, 6b, unfrozen)}} & \gc{6.24 kbps} & \gc{1020} & \gc{\textbf{4.372}} & \gc{\textbf{3.356}} & \gc{\textbf{0.975}} & \gc{0.944} \\
    \cmidrule(l){2-9}
    & \multirow{8}{*}{\rotatebox{90}{\begin{tabular}{@{}c@{}}\textbf{3-5\,kbps} \\ mid\end{tabular}}} & DAC & 4.0 kbps & 400 & 3.811 & 2.762 & 0.934 & \textbf{0.952} \\
    & & HiFi-Codec & 4.0 kbps & 400 & 4.166 & 2.763 & 0.945 & 0.950 \\
    & & HiFi-Codec & 3.0 kbps & 300 & 4.269 & 2.909 & 0.949 & 0.947 \\
    & & EnCodec & 3.0 kbps & 300 & 2.391 & 2.019 & 0.906 & 0.933 \\
    & & Vocos & 3.0 kbps & 300 & 3.788 & 2.501 & 0.931 & 0.939 \\
    & & SpeechTokenizer & 3.0 kbps & 300 & 3.991 & 2.046 & 0.902 & 0.930 \\
    & & \gc{GS-Codec ($N_G{=}80$, 4b, frozen)} & \gc{3.52 kbps} & \gc{880} & \gc{4.256} & \gc{2.998} & \gc{0.960} & \gc{0.942} \\
    & & \gc{\textbf{GS-Codec ($N_G{=}80$, 4b, unfrozen)}} & \gc{3.79 kbps} & \gc{907} & \gc{\textbf{4.338}} & \gc{\textbf{3.167}} & \gc{\textbf{0.968}} & \gc{0.944} \\
    \cmidrule(l){2-9}
    & \multirow{5}{*}{\rotatebox{90}{\begin{tabular}{@{}c@{}}\textbf{${<}$3\,kbps} \\ low\end{tabular}}} & DAC & 1.0 kbps & 100 & 1.444 & 1.208 & 0.782 & 0.810 \\
    & & WavTokenizer & 0.5 kbps & 40 & 4.019 & 2.114 & 0.909 & 0.941 \\
    & & WavTokenizer & 0.9 kbps & 75 & \textbf{4.258} & 2.492 & 0.931 & 0.940 \\
    & & \gc{GS-Codec ($N_G{=}60$, 4b, frozen)} & \gc{2.64 kbps} & \gc{660} & \gc{3.602} & \gc{2.312} & \gc{0.925} & \gc{0.929} \\
    & & \gc{\textbf{GS-Codec ($N_G{=}60$, 4b, unfrozen)}} & \gc{2.84 kbps} & \gc{680} & \gc{4.046} & \gc{\textbf{2.666}} & \gc{\textbf{0.950}} & \gc{\textbf{0.941}} \\
    \bottomrule
  \end{tabularx}
\end{table}

\subsubsection{Comparison with open-source codecs}
\label{sec:comparison}

Table~\ref{tab:wavtokenizer_comparison} presents reconstruction results on LibriTTS \texttt{test-clean}, \texttt{test-other}, and LJSpeech.
At the high-bitrate operating point ($N_G{=}90$, 6b, unfrozen), GS-Codec surpasses all comparable-bitrate baselines on UTMOS, PESQ, and STOI, approaching the DAC 9\,kbps reference.
The trend holds across all three evaluation sets and degrades gracefully at lower rates ($N_G{=}60$, 4b, unfrozen, 2.84\,kbps still leads multi-codebook baselines at 3\,kbps on UTMOS).
WavTokenizer remains stronger below 1\,kbps, outside our operating range.

Table~\ref{tab:comparison_baselines} compares GS-Codec trained on the larger Librilight-Large corpus (54k\,h) against additional baselines on LibriSpeech \texttt{test-clean}, adding WER and SIM.
At 6.05\,kbps GS-Codec leads on UTMOS (4.01) and SIM (0.94), while DAC retains a NB-PESQ advantage (3.77 vs.\ 3.71).

\begin{table}[ht]
 \caption{Comparison with open-source neural audio codecs on LibriSpeech \texttt{test-clean}.
  WER evaluated with \texttt{facebook/hubert-large-ls960-ft} on 4-10\,s subset.
  Acoustic metrics on RVQ1:8 for stacked codecs; for GS-Codec this is the full 33-row token
  stream (see Eq.~\ref{eq:bitrate}). Baseline results from published results.}
  \label{tab:comparison_baselines}
  \centering
  \small
  \setlength{\tabcolsep}{3pt}
  \scalebox{0.85}{
  \begin{tabularx}{1.111\textwidth}{c Xcccccc}
    \toprule
    & \textbf{Model} & \textbf{BR} & \textbf{Param}
    & \textbf{WER}$\downarrow$
    & \textbf{PESQ}$\uparrow$ & \textbf{UTMOS}$\uparrow$
    & \textbf{SIM}$\uparrow$ \\
    & & (kbps) & & (\%) & & & \\
    \midrule
    \multirow{5}{*}{\rotatebox{90}{\scriptsize \begin{tabular}{@{}c@{}}\textbf{$\geq$5\,kbps} \\ high\end{tabular}}}
    & DAC-75Hz           & 6.0  & 74M  & 2.27 & \textbf{3.77} & 3.62 & 0.90 \\
    & EnCodec-75Hz       & 6.0  & 15M  & 2.24 & 3.12 & 3.01 & 0.89 \\
    & \gc{GS-Codec (130g, 4b, frozen)} & \gc{5.72} & \gc{55M} & \gc{2.25} & \gc{3.67} & \gc{3.93} & \gc{0.93} \\
    & \gc{GS-Codec (120g, 4b, frozen)} & \gc{5.28} & \gc{55M} & \gc{2.22} & \gc{3.65} & \gc{3.94} & \gc{0.93} \\
    & \gc{GS-Codec (110g, 5b, frozen)} & \gc{6.05} & \gc{55M} & \gc{\textbf{2.19}} & \gc{3.71} & \gc{\textbf{4.01}} & \gc{\textbf{0.94}} \\
    \cmidrule(l){1-8}
    \multirow{4}{*}{\rotatebox{90}{\scriptsize \begin{tabular}{@{}c@{}}\textbf{3-5\,kbps} \\ mid\end{tabular}}}
    & SpeechTokenizer-50Hz & 4.0 & 103M & 2.47 & 3.01 & 3.90 & 0.85 \\
    & \gc{GS-Codec (70g, 5b, frozen)} & \gc{3.85} & \gc{55M} & \gc{2.29} & \gc{3.41} & \gc{3.88} & \gc{0.91} \\
    & \gc{GS-Codec (80g, 4b, frozen)} & \gc{3.52} & \gc{55M} & \gc{2.27} & \gc{3.46} & \gc{3.91} & \gc{0.91} \\
    & \gc{GS-Codec (110g, 4b, frozen)} & \gc{4.84} & \gc{55M} & \gc{\textbf{2.26}} & \gc{\textbf{3.63}} & \gc{\textbf{3.95}} & \gc{\textbf{0.93}} \\
    \cmidrule(l){1-8}
    \multirow{3}{*}{\rotatebox{90}{\scriptsize \begin{tabular}{@{}c@{}}\textbf{${<}$3\,kbps} \\ low\end{tabular}}}
    & Mimi-12.5Hz        & 1.1 & 78M  & 3.15 & 2.75 & 3.56 & 0.73 \\
    & XYTokenizer-12.5Hz & 1.0 & 520M & \textbf{2.36} & 3.00 & \textbf{4.00} & 0.84 \\
    & \gc{GS-Codec (60g, 4b, unfrozen)} & \gc{2.84} & \gc{55M} & \gc{2.43} & \gc{\textbf{3.02}} & \gc{3.54} & \gc{\textbf{0.87}} \\
    \bottomrule
  \end{tabularx}
  }
\end{table}

Table~\ref{tab:controlled} compares GS, FSQ, and RVQ using the same 32-channel, 64-filter SEANet backbone and 1,237-utterance LibriSpeech \texttt{test-clean} subset. GS achieves higher UTMOS and SIM, with comparable PESQ-NB, STOI, and WER at the reported precision. Appendix~\ref{app:controlled} gives the training and evaluation settings.

\begin{table}[ht]
\centering
\caption{Same-backbone comparison at approximately 6\,kbps. Values are means $\pm$ 95\% utterance-bootstrap confidence interval half-widths ($n=1{,}237$). PESQ is narrowband. GS uses predicted centers.}
\label{tab:controlled}
\small
\setlength{\tabcolsep}{4pt}
\begin{tabular}{lccc}
\toprule
 & RVQ & FSQ & GS Predictor \\
\midrule
Payload (kbps) & 6.000 & 6.000 & 5.950 \\
PESQ-NB $\uparrow$ & $3.8\pm.011$ & $3.8\pm.013$ & $3.8\pm.009$ \\
UTMOS $\uparrow$ & $3.94\pm.020$ & $3.94\pm.019$ & $4.05\pm.017$ \\
STOI $\uparrow$ & $.96\pm.0008$ & $.96\pm.0008$ & $.96\pm.0011$ \\
SIM $\uparrow$ & $.93\pm.0013$ & $.92\pm.0014$ & $.94\pm.0012$ \\
WER (\%) $\downarrow$ & $2.1\pm.25$ & $2.2\pm.26$ & $2.1\pm.25$ \\
\bottomrule
\end{tabular}
\end{table}

\paragraph{Human listening evaluation.}
Blinded MUSHRA and MOS studies at approximately 3 and 6\,kbps compare GS, FSQ, and RVQ, with DAC and EnCodec as external reference points. At approximately 6\,kbps GS obtains higher mean MUSHRA scores and smaller mean MOS gaps to natural speech than FSQ and RVQ; at approximately 3\,kbps it obtains smaller mean MOS gaps at similar MUSHRA scores. Appendix~\ref{app:listening} gives the protocol and all scores. Audio samples are available at \url{https://ronaluf.github.io/gs-codec/}.

\subsubsection{Bitrate sweep}
\label{sec:sweep}

\begin{figure}[ht]
  \centering
  \includegraphics[width=0.85\linewidth]{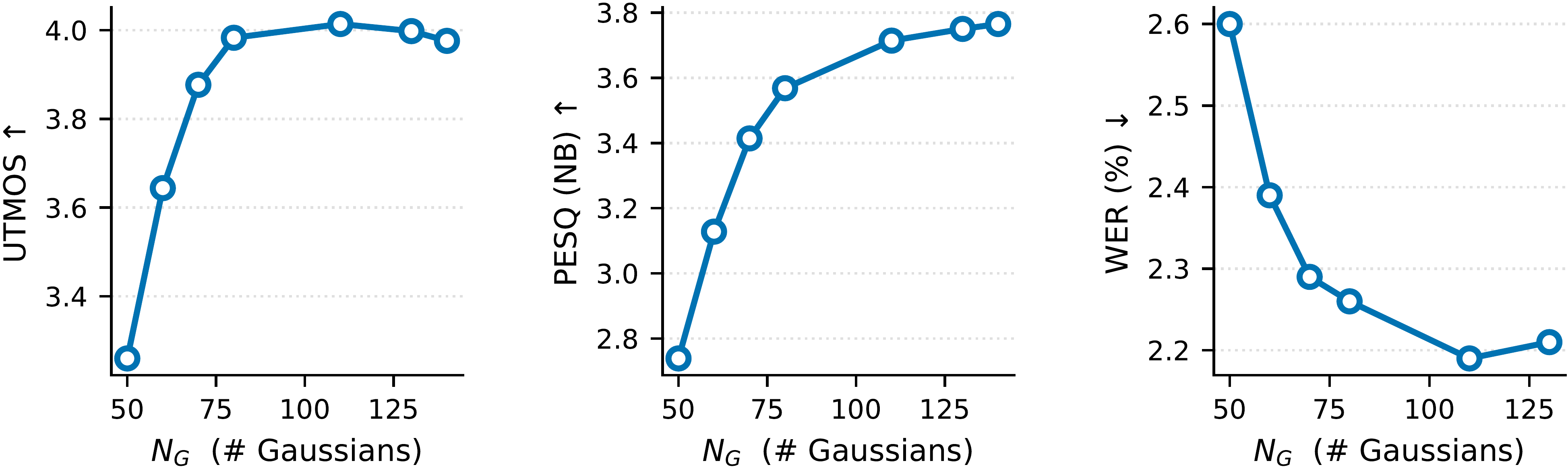}
  \caption{$N_G$ sweep at fixed 5-bit PTQ on LibriSpeech \texttt{test-clean}
    (4-10\,s subset, $n{=}1237$; WER: \texttt{hubert-large-ls960-ft}). }
  \label{fig:ng_ablation_5bit}
\end{figure}

\begin{figure}[ht]
  \centering
  \includegraphics[width=0.8\linewidth]{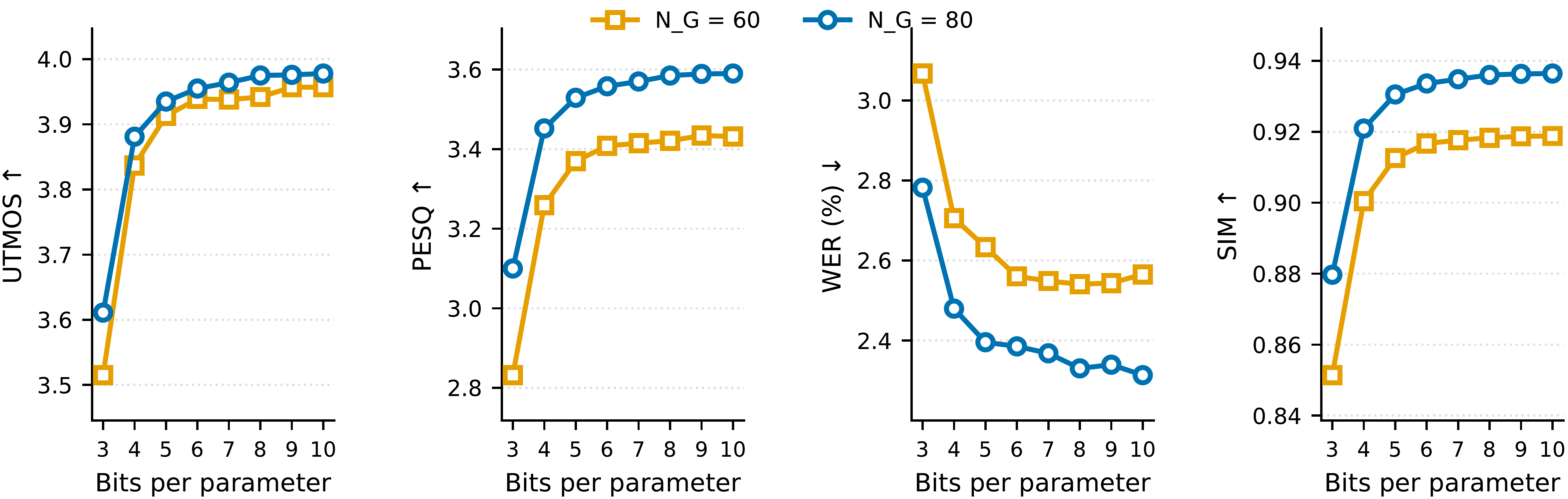}
  \caption{Bit-depth sweep for $N_G{=}60$ (orange squares) and $N_G{=}80$
    (blue circles) on LibriSpeech \texttt{test-clean} (4-10\,s subset, $n{=}1237$).
    Metrics saturate past 6\,bits; $N_G{=}80$ outperforms $N_G{=}60$ at every bit-depth.}
  \label{fig:ng60_vs_80_bits_sweep}
\end{figure}

Figure~\ref{fig:ng_ablation_5bit} sweeps $N_G \in \{50,\ldots,140\}$ at fixed 5-bit depth (2.75-7.70\,kbps).
All three metrics improve steeply up to $N_G{=}80$ and saturate thereafter.
Figure~\ref{fig:ng60_vs_80_bits_sweep} sweeps bit-depth $B\in\{3,\ldots,10\}$ for $N_G{=}60$ and $N_G{=}80$; metrics saturate past 6\,bits while $N_G{=}80$ dominates at every bit-depth.

\subsubsection{Ablations}
\label{sec:ablations}

\begin{figure}[ht]
  \centering
  \includegraphics[width=0.7\linewidth]{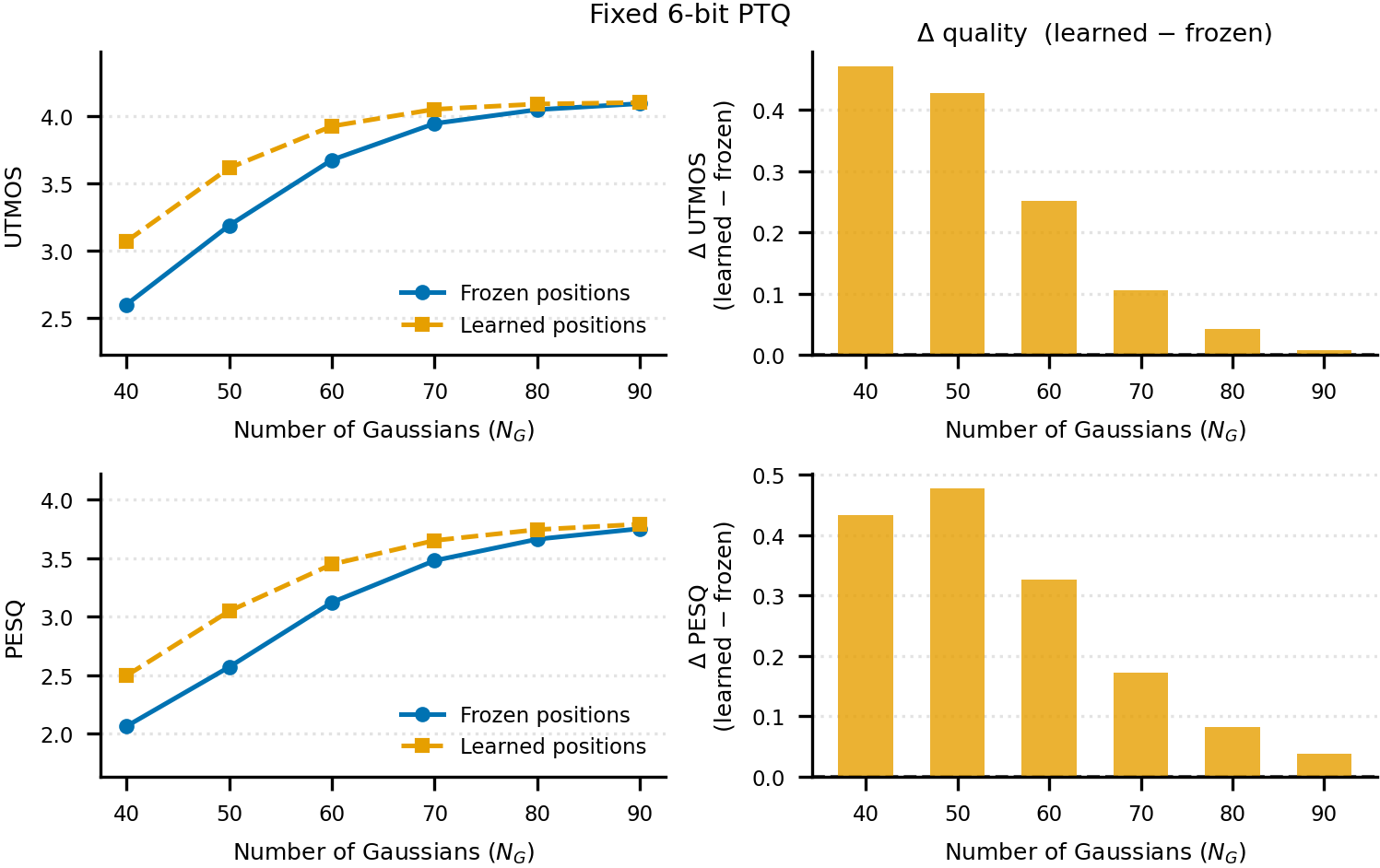}
  \caption{Effect of unfreezing primitive positions at fixed 6-bit PTQ\@.
    Left column: absolute UTMOS and PESQ versus $N_G$ for frozen (solid blue)
    and learned (dashed orange) positions. Right column: metric delta
    $\Delta = \text{learned} - \text{frozen}$.}
  \label{fig:unfrozen_comparison}
\end{figure}

We ablate the design choice of fixing primitive positions to a uniform temporal
grid versus allowing them to be optimized freely during the inner loop. This
choice represents a critical bitrate optimization, as the additional bits
required to transmit learned positions do not translate to large perceptual gains at
sufficient primitive counts. Figure~\ref{fig:unfrozen_comparison} reports
absolute quality and the metric delta between the two conditions across all
swept values of $N_G$ at fixed 6-bit post-training quantization.

\subsection{Encoding time and throughput}
\label{sec:latency}

Table~\ref{tab:latency} reports per-segment GPU execution time on an NVIDIA L40S GPU
(200 segments $\times$ 3\,s, 10 warmup segments, measured with \texttt{torch.cuda.Event}).
The iterative GS encoder (300-step Adam, paper evaluation config) takes 171\,ms per segment
(RTF\,=\,0.057), roughly $34\times$ slower than DAC and $32\times$ slower than EnCodec.
The GS Predictor Net reduces this to 7.9\,ms
(RTF\,=\,0.003), competitive with DAC ($1.6\times$) and EnCodec ($1.5\times$). Measurements exclude the decoder; the timed predictor uses frozen centers.

\begin{table}[ht]
  \caption{Encoder GPU execution time on NVIDIA L40S. RTF $= t_{\text{encode}} / t_{\text{audio}}$;
  lower is better. Measured over 200 segments $\times$ 3\,s with 10 warmup segments.
  \texttt{torch.cuda.Event} timing; single run.}
  \label{tab:latency}
  \centering
  \small
  \begin{tabular}{lrrrr}
    \toprule
    \textbf{Model} & \textbf{Mean} & \textbf{Median} & \textbf{p95} & \textbf{RTF} \\
    \midrule
    GS-Codec iterative (300-step Adam) & 171.2\,ms & 152.6\,ms & 208.4\,ms & 0.057 \\
    GS-Pred                            &   7.9\,ms &   7.9\,ms &   7.9\,ms & 0.0026 \\
    EnCodec 24\,kHz @ 6\,kbps          &   5.4\,ms &   5.3\,ms &   5.4\,ms & 0.0018 \\
    DAC 24\,kHz @ 6\,kbps              &   5.0\,ms &   5.0\,ms &   5.1\,ms & 0.0017 \\
    \bottomrule
  \end{tabular}
\end{table}

\section{Conclusion}
\label{sec:conclusion}

We introduced GS-Codec, a neural audio codec that replaces conventional quantization bottlenecks with a parametric one-dimensional Gaussian-splatting decomposition. Instead of relying on residual vector quantization or finite scalar quantization, GS-Codec represents encoder latents as a weighted sum of Gaussian primitives, enabling compression through structured signal decomposition.

Experimental results demonstrate that GS-Codec achieves competitive or superior perceptual quality, intelligibility, and speaker similarity compared to established open-source codecs across multiple bitrate regimes. In addition, the proposed GS Predictor Net amortizes the iterative fitting process into a single forward pass, substantially reducing encoding time while maintaining reconstruction quality close to the full optimization-based approach.

Overall, our results suggest that parametric latent decompositions provide a promising alternative to partition-based neural audio compression and open new directions for fine-grained post-training rate control in neural codecs.

{\small
\bibliographystyle{plainnat}
\bibliography{references}
}
\clearpage

\appendix

\section{Training and evaluation details}
\label{app:details}

\paragraph{SEANet architecture at 24\,kHz.}
The encoder and decoder use the SEANet convolutional architecture~\citep{soundstream}
with latent dimension $C{=}32$, $n_{\mathrm{filters}}{=}64$ base channels, Snake
activation~\citep{snake}, strided ratios $[8,5,4,2]$ (total downsampling factor 320),
kernel size 7, residual kernel size 3, 1 residual layer per block, dilation base 2,
2 LSTM layers, weight-norm, and non-causal (bidirectional) convolutions.
This yields a 75\,Hz latent frame rate and $T{=}225$ frames per 3-second segment.

\paragraph{GS bottleneck hyperparameters.}
\begin{center}
\small
\begin{tabular}{ll}
\toprule
\textbf{Parameter} & \textbf{Value} \\
\midrule
Basis type & Gaussian \\
$N_G$ (training) & 80 \\
$K$ inner steps (training) & 100 \\
Inner LR (training) & 0.4 \\
Inner LR schedule (training) & none \\
$K$ inner steps (evaluation) & 300 \\
Inner LR (evaluation) & 0.05 \\
Inner LR schedule (evaluation) & cosine (5-step warmup, final ratio 0.01) \\
Position LR multiplier & 1.0 (training) / 2.0 (evaluation) \\
Positions & uniform grid $\mu_i = (i-1)(T-1)/(N_G-1)$ \\
Bias & frozen to zero \\
Scales & shared across channels \\
Amplitudes & per-channel \\
Bound parameters & false (explicit clamping at eval: $\sigma \in [0.1, 100]$, $w \in [-3, 3]$) \\
Input normalisation & disabled \\
\bottomrule
\end{tabular}
\end{center}

\paragraph{Loss weights and balancer.}
L1 time-domain: 0.1; multi-scale spectrogram: 2.0; adversarial (MS-STFT discriminator): 4.0;
feature matching: 4.0; commitment penalty: 20.0; SI-SNR: 0 (disabled).
Gradient balancer: \texttt{balance\_grads=true}, EMA decay 0.999, per-batch-item normalisation,
total norm target 1.0. Discriminator: single MS-STFT discriminator with FFT sizes
$[1024, 2048, 512, 256, 128]$ and corresponding hop lengths $[256, 512, 128, 64, 32]$.

\paragraph{Outer optimiser.}
Adam, $\beta_1{=}0.5$, $\beta_2{=}0.9$, no weight decay, LR $10^{-4}$, no LR scheduler.
EMA of model weights: decay 0.99, updated every step, stored on GPU.
Training dtype: bfloat16 (autocast disabled; bfloat16 used for gradient accumulation stability).

\paragraph{Dataset and training schedule.}
200 epochs $\times$ 2{,}000 gradient steps; 500{,}000 samples drawn per epoch with
replacement, batch size 16, 8 data workers. Segments of 3 seconds; shorter clips
accepted if $\geq 0.5 \times$ segment duration. Trained on a single NVIDIA RTX PRO 6000 GPU ($\approx$144 hours).

\paragraph{Evaluation commands.}
WER: \texttt{facebook/hubert-large-ls960-ft}, 4-10\,s duration filter on LibriSpeech
\texttt{test-clean} ($n{=}1{,}237$ files). PESQ: narrowband (NB) for Table~\ref{tab:comparison_baselines}; wideband (WB) for WavTokenizer comparison
(Table~\ref{tab:wavtokenizer_comparison}). UTMOS: \texttt{sarulab-speech/utmos22\_strong}.
SIM: WavLM-Large speaker verification cosine similarity~\citep{wavlm}.

\paragraph{GS Predictor Net training.}
The predictor is trained with the codec encoder and decoder frozen, using only
the splat-reconstruction penalty loss ($\lambda_{\mathrm{penalty}}{=}1.0$; all
other losses disabled). Architecture: $d_{\mathrm{model}}{=}128$, 4 attention
heads, 4 transformer blocks, cross-attention from $N_G$ learned queries into
dilated-conv features of $Z$. Training: Adam, LR $5{\times}10^{-4}$, cosine
schedule (1{,}000-step warmup, min ratio 0.1), batch size 16, 50 epochs
$\times$ 2{,}000 steps, 3-second segments, no refinement iterations
($K_{\mathrm{refine}}{=}0$), no parameter noise. The predictor is initialised
from the codec checkpoint and trained on the same data mixture.

\section{Per-parameter quantisation ranges}
\label{app:ptq_ranges}

Table~\ref{tab:ptq_ranges} reports the per-type quantisation ranges used for the
6-bit $k$-means PTQ codebooks shipped with the headline checkpoint.
Ranges are estimated once on 30 held-out files from \texttt{libritts-dev-clean} and
stored alongside the checkpoint as \texttt{global\_ranges\_qat.json}.

\begin{table}[H]
  \centering
  \caption{Per-parameter $k$-means PTQ codebooks (6-bit, $2^6{=}64$ centroids).
    Ranges estimated on held-out calibration set; loaded at dequantisation.}
  \label{tab:ptq_ranges}
  \small
  \begin{tabular}{lrrr}
    \toprule
    \textbf{Parameter} & \textbf{Min} & \textbf{Max} & \textbf{Levels ($2^B$)} \\
    \midrule
    Scale $\sigma_i$ & 0.5 & 56.25 & 64 \\
    Amplitude $w_{i,c}$ & $-3.0$ & $3.0$ & 64 \\
    \bottomrule
  \end{tabular}
\end{table}

\section{Controlled evaluation}
\label{app:controlled}

\paragraph{Setup.}
RVQ, FSQ, and GS share the SEANet encoder/decoder ($C{=}32$, 64 filters, Snake), Librilight-Large, and a 200-epoch $\times$ 2,000-step schedule. Encoder/decoder: 55.45M parameters. Bottleneck state (parameters and buffers): FSQ 32, GS iterative 0, RVQ 532,488, GS Predictor 1,076,998. RVQ: $n_q{=}8\times1{,}024$. FSQ: $[4]{\times}16+[8]{\times}16$, 80 bits/frame, 6.00\,kbps. GS: 102 primitives, 5-bit scales and amplitudes, 10-bit centers, $102(33\times5+10)=17{,}850$ bits per 3\,s, 5.950\,kbps. Evaluation: LibriSpeech \texttt{test-clean}, 4--10\,s, pinned 1,237 utterances, PESQ-NB, WER with \texttt{hubert-large-ls960-ft}. Intervals are paired 95\% bootstrap half-widths over utterances; pairwise tests are Wilcoxon signed-rank with Holm correction. Token rate counts one token per quantized scalar or codebook index: GS has the lowest temporal rate but $1.93\times$ the token rate of RVQ and $0.48\times$ that of FSQ (Table~\ref{tab:discussion_extra}).

\begin{table}[ht]
\centering
\caption{Unified evaluation on the pinned 1,237 utterances. DAC and EnCodec are public checkpoints rerun with the same protocol.}
\label{tab:discussion_extra}
\small
\setlength{\tabcolsep}{3.5pt}
\resizebox{\linewidth}{!}{\begin{tabular}{lrccccccrr}
\toprule
System & kbps & PESQ-NB $\uparrow$ & UTMOS $\uparrow$ & STOI $\uparrow$ & SIM $\uparrow$ & WER $\downarrow$ & Steps/s & Tok./step & Tok./s \\
\midrule
GS iterative, 102g/5b & 5.95 & $3.8\pm.009$ & $4.07\pm.017$ & $.96\pm.0010$ & $.94\pm.0012$ & $2.1\pm.26$ & 34 & 34 & 1,156 \\
GS Predictor, 102g/5b & 5.95 & $3.8\pm.009$ & $4.05\pm.017$ & $.96\pm.0011$ & $.94\pm.0012$ & $2.1\pm.25$ & 34 & 34 & 1,156 \\
FSQ, 32-dim & 6.00 & $3.8\pm.013$ & $3.94\pm.019$ & $.96\pm.0008$ & $.92\pm.0014$ & $2.2\pm.26$ & 75 & 32 & 2,400 \\
RVQ, 8 codebooks & 6.00 & $3.8\pm.011$ & $3.94\pm.020$ & $.96\pm.0008$ & $.93\pm.0013$ & $2.1\pm.25$ & 75 & 8 & 600 \\
DAC-75Hz & 6.00 & $3.7\pm.010$ & $3.62\pm.023$ & $.95\pm.0007$ & $.88\pm.0017$ & $2.3\pm.26$ & 75 & 8 & 600 \\
EnCodec-75Hz & 6.00 & $3.1\pm.018$ & $3.09\pm.027$ & $.93\pm.0010$ & $.89\pm.0015$ & $2.2\pm.26$ & 75 & 8 & 600 \\
\bottomrule
\end{tabular}}
\end{table}

\paragraph{Subgroup analysis.}
Same 1,237 utterances; for each condition the extreme tercile ($n{=}412$) by median voiced F0, a ZCR/high-band-energy frication proxy, 95th-percentile spectral flux (transients), fraction of frames below 5\% peak RMS (low energy), and peak-to-floor RMS ratio (background level). GS Predictor (102g/5b, 5.950\,kbps) against same-backbone FSQ and RVQ (6.000\,kbps); paired bootstrap CIs and Wilcoxon tests with Holm correction (Table~\ref{tab:subgroups}).

\begin{table}[ht]
\centering
\caption{UTMOS by subgroup.}
\label{tab:subgroups}
\small
\begin{tabular}{lcccll}
\toprule
Subset & GS Predictor & FSQ & RVQ & $\Delta$ vs FSQ [95\% CI] & Holm $p$ \\
\midrule
High F0 & 3.97 & 3.84 & 3.85 & $+0.125$ [$+0.109$, $+0.140$] & $3\times10^{-43}$ \\
High frication proxy & 4.01 & 3.88 & 3.90 & $+0.122$ [$+0.108$, $+0.137$] & $6\times10^{-47}$ \\
High spectral flux & 4.04 & 3.95 & 3.95 & $+0.096$ [$+0.083$, $+0.109$] & $8\times10^{-40}$ \\
Low energy & 4.05 & 3.94 & 3.93 & $+0.114$ [$+0.100$, $+0.129$] & $7\times10^{-46}$ \\
Lowest peak-to-floor ratio & 3.98 & 3.86 & 3.86 & $+0.119$ [$+0.106$, $+0.132$] & $2\times10^{-50}$ \\
All & 4.05 & 3.94 & 3.94 & $+0.109$ [$+0.101$, $+0.116$] & $2\times10^{-133}$ \\
\bottomrule
\end{tabular}
\end{table}

\paragraph{Representation ablation.}
Same trained encoder and utterances, payloads within 0.85\% of 17,850 bits, plain PTQ. Basis rows send 32 amplitudes and one width per primitive at 5 bits, plus a 10-bit center when sent; Gabor adds frequency and phase. Removing the decomposition roughly doubles NMSE; changing the atom family changes it by at most 17\% (Table~\ref{tab:basis}).

\begin{table}[ht]
\centering
\caption{Latent NMSE at matched payload.}
\label{tab:basis}
\begin{tabular}{lrrr}
\toprule
Representation & Values & Bits / 3\,s & NMSE $\downarrow$ \\
\midrule
Gaussian, free centers, 102 & 3,468 & 17,850 & .0259 \\
Triangular, 102 & 3,468 & 17,850 & .0275 \\
Laplacian, 102 & 3,468 & 17,850 & .0295 \\
Gabor, 97 & 3,492 & 17,945 & .0303 \\
Gaussian, grid centers, 109 & 3,597 & 17,985 & .0317 \\
No decomposition: decimate to 102 frames & 3,264 & 17,850 & .0496 \\
No decomposition: framewise SQ & 7,200 & 18,000 & .0537 \\
\bottomrule
\end{tabular}
\end{table}

\paragraph{Continuous-autoencoder control.}
Same backbone; the bottleneck linearly resamples the latent $225\to102$ frames and back during training (no quantizer), followed by per-channel $k$-means PTQ. Single checkpoint trained at 102 frames. At its trained count the AE trails GS by 0.066 UTMOS with fewer bits; away from it, by 0.85--0.88 UTMOS at matched bits (Table~\ref{tab:ae}).

\begin{table}[ht]
\centering
\caption{GS versus a continuous autoencoder with post-hoc $k$-means. UTMOS on the 1,237 utterances.}
\label{tab:ae}
\small
\begin{tabular}{lrrlrrr}
\toprule
GS $N_G$/bits & kbps & UTMOS & AE frames/bits & kbps & UTMOS & $\Delta$ \\
\midrule
102/5 & 5.950 & 4.051 & 102/5 (trained) & 5.440 & 3.985 & $+0.066$ \\
91/5 & 5.005 & 3.699 & 94/5 & 5.013 & 2.823 & $+0.876$ \\
131/4 & 5.764 & 3.969 & 136/4 & 5.803 & 3.118 & $+0.851$ \\
127/5 & 6.985 & 4.025 & 131/5 & 6.987 & 3.145 & $+0.880$ \\
\bottomrule
\end{tabular}
\end{table}

\paragraph{Training cost.}
Wall-clock time per optimizer update; single L40S, bf16, batch 16, identical backbone, data, and losses (Table~\ref{tab:traincost}). The inner fit is batched and its history is not traversed by the outer backward pass.

\begin{table}[ht]
\centering
\caption{Training cost per update.}
\label{tab:traincost}
\begin{tabular}{lrrr}
\toprule
Bottleneck & Time/update & Updates/s & 400k updates \\
\midrule
FSQ & 1.160\,s & 0.862 & $\approx$5.3 days \\
GS, 100-step inner loop & 1.209\,s & 0.827 & $\approx$5.6 days \\
RVQ, 8 codebooks & 1.319\,s & 0.758 & $\approx$6.1 days \\
\bottomrule
\end{tabular}
\end{table}

\FloatBarrier
\section{Human listening evaluation}
\label{app:listening}

Ten LibriSpeech \texttt{test-clean} utterances (4.0--5.5\,s), one per speaker, selected by a seeded random procedure without inspecting method scores. MOS: each condition rated independently on a 1--5 scale (``Listen to each clip once or twice, then rate its overall quality from 1 to 5''). MUSHRA: blinded 0--100 interface with a hidden reference and a 3.5-kHz anchor, randomized order (``Rate the overall audio quality of each sample relative to the reference''). Participants confirmed headphone use and a quiet room; they were unpaid volunteers and took part anonymously. Ratings are averaged within listener, then across listeners; intervals are participant-level bootstrap. MOS gap is natural speech minus system, paired within listener; lower is better.

At $\approx$6\,kbps all GS configurations obtain higher mean MUSHRA and smaller mean MOS gaps than FSQ and RVQ; at $\approx$3\,kbps they obtain smaller mean MOS gaps at similar MUSHRA (Table~\ref{tab:listening}). The Predictor's MOS gap is smaller at $\approx$3 than at $\approx$6\,kbps, with overlapping intervals; its MUSHRA ordering is as expected.

\begin{table}[ht]
\centering
\caption{Listening scores. MUSHRA: mean $\pm$ 95\% CI half-width. MOS gap: mean [95\% CI].}
\label{tab:listening}
\small
\setlength{\tabcolsep}{3pt}
\begin{tabular}{lrrl}
\toprule
System & kbps & MUSHRA $\uparrow$ & MOS gap $\downarrow$ \\
\midrule
Hidden reference / natural speech & n/a & $78.62\pm5.41$ & 0.00 [0.00, 0.00] \\
GS iterative, unfrozen, 100g/5b & 5.833 & $77.67\pm5.01$ & 0.14 [$-0.28$, 0.56] \\
GS Predictor, 109g/5b & 5.995 & $75.36\pm5.30$ & 0.61 [0.19, 1.03] \\
GS iterative, frozen, 110g/5b & 6.050 & $75.13\pm5.14$ & 0.50 [0.00, 0.97] \\
RVQ & 6.000 & $72.53\pm5.66$ & 0.69 [0.25, 1.14] \\
FSQ & 3.000 & $70.78\pm5.78$ & 0.78 [0.22, 1.31] \\
RVQ & 3.000 & $70.60\pm5.51$ & 0.78 [0.36, 1.19] \\
GS iterative, unfrozen, 63g/4b & 2.982 & $70.60\pm6.12$ & 0.39 [$-0.14$, 0.92] \\
GS iterative, frozen, 68g/4b & 2.992 & $69.49\pm5.92$ & 0.50 [0.14, 0.86] \\
GS Predictor, 80g/3b & 2.907 & $68.82\pm5.86$ & 0.33 [$-0.11$, 0.78] \\
FSQ & 6.000 & $67.74\pm6.01$ & 0.78 [0.31, 1.25] \\
DAC 24\,kHz & 6.000 & $64.22\pm5.53$ & 0.97 [0.58, 1.36] \\
EnCodec 24\,kHz & 6.000 & $53.68\pm6.25$ & 1.36 [0.92, 1.78] \\
3.5-kHz anchor & n/a & $36.19\pm5.74$ & n/a \\
\bottomrule
\end{tabular}
\end{table}

\FloatBarrier
\section{Fixed-budget allocation}
\label{app:allocation}

Frozen positions; each budget holds $N_G B$ fixed. Intervals are 95\% bootstrap half-widths, 10,000 resamples. At 5.28\,kbps, 120g/4b beats 160g/3b and 80g/6b on PESQ-NB, UTMOS, and SIM; at $N_G{=}60$, raising precision from 5 to 10 bits moves UTMOS only from 2.845 to 2.897 (Table~\ref{tab:allocation}). Sufficient primitive capacity comes first; the remaining budget goes to precision.

\begin{table}[ht]
\centering
\caption{Fixed-budget allocation between primitive count and precision. WER in percent.}
\label{tab:allocation}
\small
\setlength{\tabcolsep}{4pt}
\begin{tabular}{rrrcccc}
\toprule
kbps & $N_G$ & Bits & PESQ-NB $\uparrow$ & UTMOS $\uparrow$ & WER $\downarrow$ & SIM $\uparrow$ \\
\midrule
3.30 & 100 & 3 & $2.926\pm.055$ & $3.425\pm.098$ & $2.48\pm.74$ & $.844\pm.006$ \\
3.30 & 60 & 5 & $2.484\pm.042$ & $2.845\pm.066$ & $2.38\pm.72$ & $.797\pm.009$ \\
3.96 & 120 & 3 & $3.155\pm.051$ & $3.583\pm.103$ & $2.33\pm.73$ & $.867\pm.007$ \\
3.96 & 60 & 6 & $2.515\pm.044$ & $2.880\pm.066$ & $2.48\pm.74$ & $.804\pm.009$ \\
3.96 & 40 & 9 & $1.898\pm.031$ & $1.905\pm.065$ & $5.48\pm1.35$ & $.612\pm.013$ \\
4.40 & 100 & 4 & $3.494\pm.044$ & $3.889\pm.087$ & $2.02\pm.68$ & $.912\pm.004$ \\
4.40 & 80 & 5 & $3.181\pm.050$ & $3.593\pm.080$ & $2.33\pm.71$ & $.895\pm.005$ \\
4.40 & 40 & 10 & $1.899\pm.031$ & $1.908\pm.064$ & $5.59\pm1.35$ & $.612\pm.013$ \\
5.28 & 160 & 3 & $3.327\pm.051$ & $3.610\pm.105$ & $2.02\pm.66$ & $.889\pm.005$ \\
5.28 & 120 & 4 & $3.685\pm.038$ & $3.953\pm.086$ & $2.02\pm.66$ & $.930\pm.004$ \\
5.28 & 80 & 6 & $3.225\pm.050$ & $3.640\pm.081$ & $2.28\pm.71$ & $.901\pm.005$ \\
5.28 & 60 & 8 & $2.532\pm.044$ & $2.895\pm.068$ & $2.48\pm.74$ & $.807\pm.008$ \\
6.60 & 120 & 5 & $3.805\pm.035$ & $4.039\pm.080$ & $2.17\pm.68$ & $.944\pm.003$ \\
6.60 & 100 & 6 & $3.672\pm.039$ & $3.996\pm.083$ & $2.17\pm.69$ & $.933\pm.004$ \\
6.60 & 60 & 10 & $2.532\pm.044$ & $2.897\pm.069$ & $2.53\pm.76$ & $.807\pm.008$ \\
7.92 & 120 & 6 & $3.836\pm.034$ & $4.061\pm.078$ & $2.12\pm.67$ & $.948\pm.003$ \\
7.92 & 80 & 9 & $3.235\pm.051$ & $3.646\pm.080$ & $2.33\pm.74$ & $.903\pm.005$ \\
\bottomrule
\end{tabular}
\end{table}

\section{Limitations}
\label{app:limitations}

GS-Codec has the following limitations. First, encoding is slower than feed-forward codecs: even with the GS Predictor Net, encoding time remains higher than EnCodec and DAC, and the iterative inner loop is substantially slower when the predictor is not used. Second, at lower bitrates (Table~\ref{tab:comparison_baselines}, ${<}$3\,kbps band), perceptual quality leaves room for improvement; the baselines in this band operate at different rates. Our experiments evaluate English speech at moderate bitrates. Generalisation to other audio domains and sub-kbps coding remains future work. The current codec processes complete 3-second segments with non-causal components, so GPU execution time does not represent streaming latency.

\section{Broader impact}
\label{app:impact}
GS-Codec can reduce the storage and transmission costs of speech. As with other speech codecs, reconstructed audio can retain speaker identity and private information; use should respect data consent and licenses. Downstream speech-synthesis applications share the potential misuse risks of existing audio representations.

\end{document}